\documentclass[useAMS,natbib]{mn2e}

\usepackage{natbibmnfix, graphicx, times}
\usepackage{subfigure}
\usepackage{color,hyperref}
\usepackage{amssymb,amsmath}
\usepackage{bm}
\def\red#1 {\textcolor{red}{#1}\ }   
\def\blue#1 {\textcolor{blue}{#1}\ }   

\usepackage[english]{babel}
\usepackage{textgreek}
\usepackage{commath}
\usepackage{multirow}
\usepackage{color}
\usepackage{aecompl}
\usepackage{booktabs}
\usepackage[colorinlistoftodos]{todonotes}
\usepackage{tikz}

\newcommand{\be}{\begin{equation}}
\newcommand{\ee}{\end{equation}}
\def\ba{\begin{eqnarray}}
\def\ea{\end{eqnarray}}

\newcommand{\bk}{{\bf k}}
\newcommand{\bB}{{\bf B}}
\newcommand{\der}{{\rm d}}

\title[ IXPE Detection of Polarized X-rays from Magnetars and Photon
  Mode Conversion at QED Vacuum Resonance] {\Huge\bf
  \fontfamily{phv}\selectfont IXPE Detection of Polarized X-rays from
  Magnetars and Photon Mode Conversion at QED Vacuum Resonance}

\author[Lai]{ \bf \Large \fontfamily{phv}\selectfont Dong
  Lai\thanks{E-mail:dong@astro.cornell.edu}\\ \fontfamily{phv}\selectfont
  Cornell Center for Astrophysics and Planetary Science, Department of
  Astronomy, Cornell University, Ithaca, NY 14853}

\begin{document}
\maketitle

\begin{abstract}
{\bf \fontfamily{phv}\selectfont The recent observations of the
    anomalous X-ray pulsars 4U 0142+61 and 1RXS J170849.0-400910 by
    the Imaging X-ray Polarimetry Explorer (IXPE) opened up a new
    avenue to study magnetars, neutron stars endowed with superstrong
    magnetic fields ($B\gtrsim 10^{14}$~G). The detected polarized
    X-rays from 4U 0142+61 exhibit a 90$^\circ$ linear polarization
    swing from low photon energies ($E\lesssim 4$~keV) to high
    energies ($E\gtrsim 5.5$~keV).  We show that this swing can be
    explained by photon polarization mode conversion at the vacuum
    resonance in the magnetar atmosphere; the resonance arises from
    the combined effects of plasma-induced birefringence and
    QED-induced vacuum birefringence in strong magnetic fields.  This
    explanation suggests that the atmosphere of 4U 0142 be composed of
    partially ionized heavy elements, and the surface magnetic field
    be comparable or less than $10^{14}$~G, consistent with the dipole
    field inferred from the measured spindown. It also implies that
    the spin axis of 4U 0142+61 is aligned with its velocity
    direction. The polarized X-rays from 1RXS J170849.0-400910 do not
    show such $90^\circ$ swing, consistent with magnetar atmospheric
    emission with $B\gtrsim 5\times 10^{14}$~G.}
\end{abstract}

\begin{keywords}
  {\fontfamily{phv}\selectfont neutron stars | x-rays | magnetic fields | polarization}
\end{keywords}



\bigskip
\noindent
{\bf Significance Statement:} The recent detections of polarized X-rays
from magnetars have opened up a new avenue to study magnetars, neutron
stars endowed with superstrong magnetic fields.  The detected
polarization signals are intriguing, and can be explained by a novel
QED vacuum resonance effect -- an effect that is analogous to MSW
neutrino oscillation that operates in the Sun and similar level-crossing phenomena 
in other areas of sciences.

\bigskip
\section*{Introduction}

Magnetars are neutron stars (NSs) whose energy outputs (even in quiescence) are dominated by
magnetic field dissipations
\cite{thompsonduncan93,kaspi17}.
Recently the NASA/ASI Imaging X-ray Polarimetry Explorer (IXPE)
\cite{weisskopf16}
reported the detection of linearly polarized x-ray emission from the
anomalous x-ray pulsar (AXP) 4U 0142+61, a magnetar with an inferred
dipole magnetic field (based the spindown rate) of $\sim 10^{14}$~G
\cite{taverna22}.
This is the first time that polarized x-rays have been
detected from any astrophysical point sources. The overall
phase-averaged linear polarization degree is $12\pm 1\%$ throughout
the IXPE band (2-8~keV). Interestingly, there is a substantial
variation of the polarization signal with energy: the polarization degree is
$14\pm 1\%$ at 2-4~keV and $41\pm 7\%$ at 5.5-8~keV, while it drops
below the detector sensitivity around 4-5~keV, where the polarization
angle swings by $\sim 90^\circ$.

Taverna et al.~\cite{taverna22} considered several possibilities to explain the
observed polarization swing, and suggested that the thermal x-rays
from 4U 0142+61 are emitted from an extended region of the condensed
neutron star (NS) surface. In this scenario, the 2-4~keV radiation is
dominated by the O-mode (polarized in the plane spanned by the local
magnetic field and the photon wave vector), while the 5.5-8~keV
radiation by the X-mode (which is orthogonal to the O-mode) because of
re-procession by resonant Compton scattering (RCS) in the
magnetosphere.

While it is pre-mature to draw any firm conclusion without detailed
modeling, the "condensed surface + RCS" scenario may be problematic for
several reasons:
(i) At the surface temperature of $T_s\simeq 5\times 10^6$~K and
$B_{14}\equiv B/(10^{14}\,{\rm G})\simeq 1$, as appropriate for 4U 0142, it is unlikely that the NS
surface is in a condensed form, even if the surface composition is Fe
\cite{medin07,potekhin13},
simply because the
cohesive energy of the Fe solid is not sufficiently large
\cite{medin06}.
(ii) A stronger surface magnetic field is possible, but the emission from a condensed Fe surface
(for typical magnetic field and photon emission directions at the surface)
is dominated by the O-mode only for photon energies less than $E_c\simeq
0.1\eta Z^{2/5}B_{14}^{1/5}\,{\rm keV}$,
where $\eta\lesssim 1$
\cite{vanadelsberg05,potekhin12}.
An unrealistically strong field ($B_{14}\gtrsim 10^5$ for
$Z=26$) would be required to make $E_c\gtrsim 4$~keV.
(iii) As acknowledged by Taverna et al.~(2022)\cite{taverna22},
the assumption that the
phase-averaged low-energy photons are dominated by the O-mode would
imply that the NS spin axis (projected in the sky plane) is orthogonal
to the proper motion direction. This is in contradiction to the
growing evidence of spin-kick alignment in pulsars
\cite{lai01,johnston05,wang06,noutsos13,janka22}.

In this paper, we show that the 90$^\circ$ linear polarization swing
observed in 4U~0142 could be naturally explained by photon mode
conversion associated the "vacuum resonance" arising from QED and
plasma birefringence in strong magnetic fields. The essential physics
of this effect was already discussed in
Ref.~\cite{laiho03a}, where it was
shown that for neutron stars with H atmospheres, thermal photons with
$E\lesssim 1$~keV are polarized orthogonal to photons with $E\gtrsim
4$~keV, provided that the NS surface magnetic field is somewhat less
than $10^{14}$~G.  The purpose of this paper is to
re-examine the mode conversion effect under more general conditions
(particularly the atmosphere composition) and to present new
semi-analytic calculations of the polarization signals for parameters
relevant to 4U~0142.  Most recently, IXPE detected polarized X-rays from
another AXP 1RXS J170849.0-400910, and found that the polarization angle
remains constant as a function of $E$
\cite{zane23}. This result is expected for the magnetar
atmospheric emission with $B\gtrsim 5\times 10^{14}$~G -- we comment on this at the end
of the paper.

\section*{Vacuum resonance and mode conversion}

Quantum electrodynamics (QED) predicts that 
in a strong magnetic field the vacuum becomes birefringent
\cite{heisenbergeuler36,schwinger51,adler71,tsaierber75,heylhernquist97}.
This vacuum birefringence is significant for $B\gg B_Q
=m_e^2c^3/(e\hbar)=4.414\times 10^{13}$~G, the critical QED
field strength. However, when combined with the 
birefringence due to the magnetized plasma, vacuum polarization can greatly
affect radiative transfer even when the field strength is modest. A ``vacuum
resonance'' arises when the contributions from the plasma and vacuum
polarization to the dielectric tensor ``compensate'' each other
\cite{gnedinetal78,meszarosventura79,pavlovgnedin84,laiho02,laiho03a,laiho03b,
holai03}.

\begin{figure}
\centering
\includegraphics[height=7.2cm]{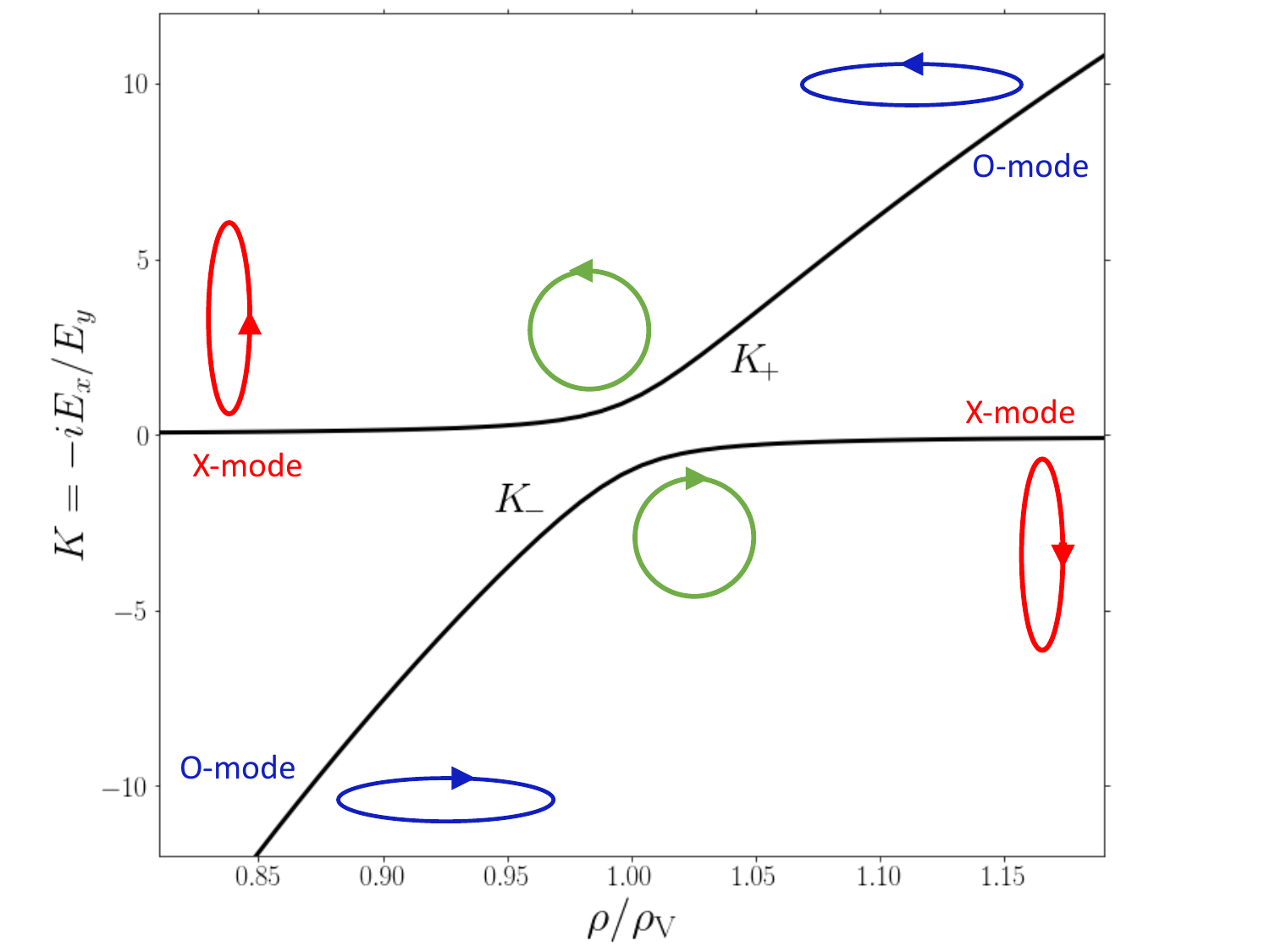}
\vskip -0cm
\caption{
The polarization ellipticity of the photon mode  
as a function of density near the vacuum resonance. 
The two curves correspond to the (+) and (-) modes.
In this example, the parameters are $B=10^{14}$G, 
$E=5\,$keV, and $\theta_{\rm kB}=30^\circ$. 
The ellipticity of a mode is specified by the ratio $K=-iE_x/E_y$, 
where $E_x$ ($E_y$) is the photon's electric field component 
along (perpendicular to) the $\bk$-$\bB$ plane. The O-mode
is characterized by $|K|\gg 1$, and the X-mode $|K|\ll 1$.}
\label{fig:f1}
\end{figure}

Consider x-ray photons propagating in the magnetized NS atmospheric plasma.
There are two polarization modes: the ordinary mode (O-mode) is mostly polarized parallel
to the $\bk$-$\bB$ plane, while the extraordinary mode (X-mode) is perpendicular to it, 
where $\bk$ is the photon wave vector and $\bB$ is the external magnetic 
field
\cite{meszaros92}.
Throughout the paper, we assume the photon energy $E$ satisfies 
$u_e=(E_{Be}/E)^2\gg 1$ and $E_{Bi}/E\ll 1$, where 
$E_{Be}=\hbar eB/(m_e c)=1158 B_{14}$~keV and
$E_{Bi}=0.63 (Z/A) B_{14}$~keV are the electron and ion cyclotron energies, respectively.
This distinction of photons modes is important since the two modes
have very different absorption and scattering opacities in the atmosphere plasma (see below):
while the O-mode opacities are similar to the zero-field values, 
the X-mode opacities are significantly reduced because the electric field
of the photon (EM wave), being orthogonal to $\bB$,
cannot effectively perturb the motion of the electron in the plasma when
$E\ll E_{Be}$.
However, the above description of O-mode and X-mode breaks down
near the ``vacuum resonance''. To be concrete, 
let us set up the $xyz$ coordinates with ${\bf k}$ along the
$z$-axis and ${\bf B}$ in the $x$-$z$ plane (such that $\hat{\bf B}\times
\hat{\bf k}=\sin\theta_{\rm kB}\hat{\bf y}$, where
$\theta_{\rm kB}$ is the angle between ${\bf k}$ and ${\bf B}$).
We write the transverse ($xy$) electric field of the photon mode as ${\bf E}\propto (iK,1)$. The mode
ellipticity $K$ is given by
\be
K_\pm = \beta\pm\sqrt{\beta^2+1},
\ee
where 
\be
\beta\simeq {u_e^{1/2}\sin^2\theta_{\rm kB}\over 2\cos\theta_{\rm kB}}\left(1-{\rho_V\over \rho}\right).
\ee
For a photon of energy $E$, the vacuum resonance density is given by 
\be
\rho_V\simeq 0.964\,Y_e^{-1}B_{14}^2E_1^2 f^{-2}~{\rm g~cm}^{-3}
\label{eq:densvp}
\ee
where $Y_e=\langle Z/A\rangle$ is the electron fraction,
$E_1=E/(1~{\rm keV})$, and $f=f(B)$ is a slowly varying function of
$B$ and is of order unity ($f=1$ for $B\ll B_Q$, $f\simeq 0.991$ at
$B_{14}=1$ and $f\rightarrow (B/5B_Q)^{1/2}$ for $B\gg B_Q$; see
\cite{potekhin04} for a general fitting formula).
For $\rho>\rho_V$ (where the plasma
effect dominates the dielectric tensor) and $\rho<\rho_V$ (where
vacuum polarization dominates), the photon modes (for typical
$\theta_{\rm kB}\neq 0$) are almost linearly polarized; near
$\rho=\rho_V$, however, the normal modes become circularly polarized
as a result of the ``cancellation'' of the plasma and vacuum effects
(see Fig.~\ref{fig:f1}). The half-width of the vacuum resonance
(defined by $|\beta|<1$) is
\begin{equation}
\epsilon\equiv {\Delta\rho\over\rho_V}={2\cos\theta_{\rm kB}\over u_e^{1/2}\sin^2\theta_{\rm kB}}.
\label{eq:width}\end{equation}

When a photon propagates in an inhomogeneous medium, its
polarization state will evolve adiabatically (i.e. following
the $K_+$ or $K_-$ curve in Fig.~\ref{fig:f1}) if the density variation is
sufficiently gentle. Thus, a X-mode (O-mode) photon will be converted
into the O-mode (X-mode) as it traverses the vacuum resonance, with its
polarization ellipse rotated by $90^\circ$ (Fig.~\ref{fig:f1}).
This resonant mode conversion is analogous to the 
Mikheyev-Smirnov-Wolfenstein neutrino oscillation that takes place in
the Sun
\cite{haxton95,bahcall03}
and similar level-crossing phenomena in other areas of sciences (e.g., Landau-Zener transition in
atomic physics, EM wave propagation in inhomogeneous media and metamaterials; see
Ref.~\cite{tracy14}).
%
For this conversion to be effective, the adiabatic condition must be satisfied
\cite{laiho02}
\be
E\gtrsim E_{\rm ad}=2.52\,\bigl(f\,\tan\theta_{\rm kB} \bigr)^{2/3}
\!\left({1\,{\rm cm}\over H_\rho}\right)^{1/3}\!{\rm keV},
\label{condition}\ee
where $H_\rho=|ds/d\ln\rho|$ is the density scale
height (evaluated at $\rho=\rho_V$) along the ray.
In general, the probability for non-adiabatic ``jump'' is
given by
\be
P_{\rm J}=\exp\left[-{\pi\over 2}\Bigl({E\over E_{\rm ad}}\Bigr)^3\right].
\ee
The mode conversion probability is $(1-P_{\rm J})$. 

\section*{Calculation of Polarized Emission}

To quantitatively compute the observed polarized X-ray emission from a
magnetic NS, it is necessary to add up emissions from all surface
patches of the star, taking account of beaming/anisotropy due to
magnetic fields and light bending due to general relativity
\cite{laiho03a,vanadelsberg06,zane06,shabaltas12,taverna20,caiazzo22}.
While this is conceptually straightforward, it necessarily involves
many uncertainties related to the unknown distributions of surface
temperature $T_s$ and magnetic field ${\bf B}$. In
addition, the atmosphere composition is unknown, the opacity data for
heavy atoms/ions for general magnetic field strengths are not
available, and atmosphere models for many surface patches (each with
different $T_s$ and ${\bf B}$) are needed. Finally, to determine the
phase-resolved lightcurve and polarization, the relative orientations
of the line of sight, spin axis and magnetic dipole axis are needed.
Given all these complexities, we present a
simplified, approximate calculation below. Our goal is to determine
under what conditions (in term of the magnetic field strength, surface
composition etc) the polarization swing can be produced.

We consider an atmosphere plasma composed of a single ionic species
(each with charge $Ze$ and mass $Am_p$) and electrons. This is of
course a simplification, as in reality the atmosphere would consist of
multiple ionic species with different ionizations.  With the equation
of state $P=\rho kT/(\mu m_p)$, hydrostatic balance implies that the
column density $y$ at density $\rho$ is given by \be y={\rho k T\over
  \mu m_p g}=0.41{\rho_1 T_6\over \mu g_2}\,{\rm g/cm}^2, \ee where
$T=10^6\,T_6$~K is the temperature, $\mu=A/(1+Z)$ is the ``molecular''
weight, $\rho_1$ is the density in 1~g/cm$^3$, and $g_2$ is the
surface gravity $g=(GM/R^2)(1-2GM/Rc^2)^{-1}$ in units of $2\times
10^{14}$ (For $M=1.4M_\odot$, $R=12$~km, we have $g_2\simeq 1.00$).
The density scale-height along a ray is \be H_\rho\simeq {kT\over \mu
  m_pg\cos\alpha} =0.41\,{T_6\over \mu g_2\cos\alpha}\,{\rm cm}, \ee
where $\alpha$ is the angle between the ray and the NS surface normal.

To simplify our calculations, we shall neglect the electron scattering
opacity and the bound-bound and bound-free opacities. The former is
generally sub-dominant compared to the free-free opacity, while the
latter are uncertain or unavailable.  The free-free opacity of a
photon mode (labeled by $i$) can be written as
\be
\kappa_i=\kappa_0\xi_i,
\label{eq:kappa}\ee
where the $B=0$ opacity is (setting the Gaunt factor to unity)
\begin{equation}
\kappa_0\simeq 9.3 \,(Z^3/A^2)\rho_1 T_6^{-1/2}E_1^{-3}{\cal G},
\end{equation}
with ${\cal G}=1-\exp(-E/kT)$. For the photon mode
${\bf E}\propto (iK,1)$, the dimensionless factor $\xi$ is given by
\cite{holai01,holai03}
\begin{equation}
\xi={K^2\sin^2\theta_{\rm kB}+(1+K^2\cos^2\theta_{\rm kB})/u_e
\over 1+K^2}.
\end{equation}
Figure 2 illustrates the behavior of $\xi_X$, $\xi_O$, $\xi_+$ and $\xi_-$ as a function of density.
We see that for typical $\theta_{\rm kB}$'s, $\xi_O\sim 1$ and $\xi_X\sim u_e^{-1}\ll 1$ except
near $\rho=\rho_V$.

\begin{figure}
\centering
\includegraphics[height=6.7cm]{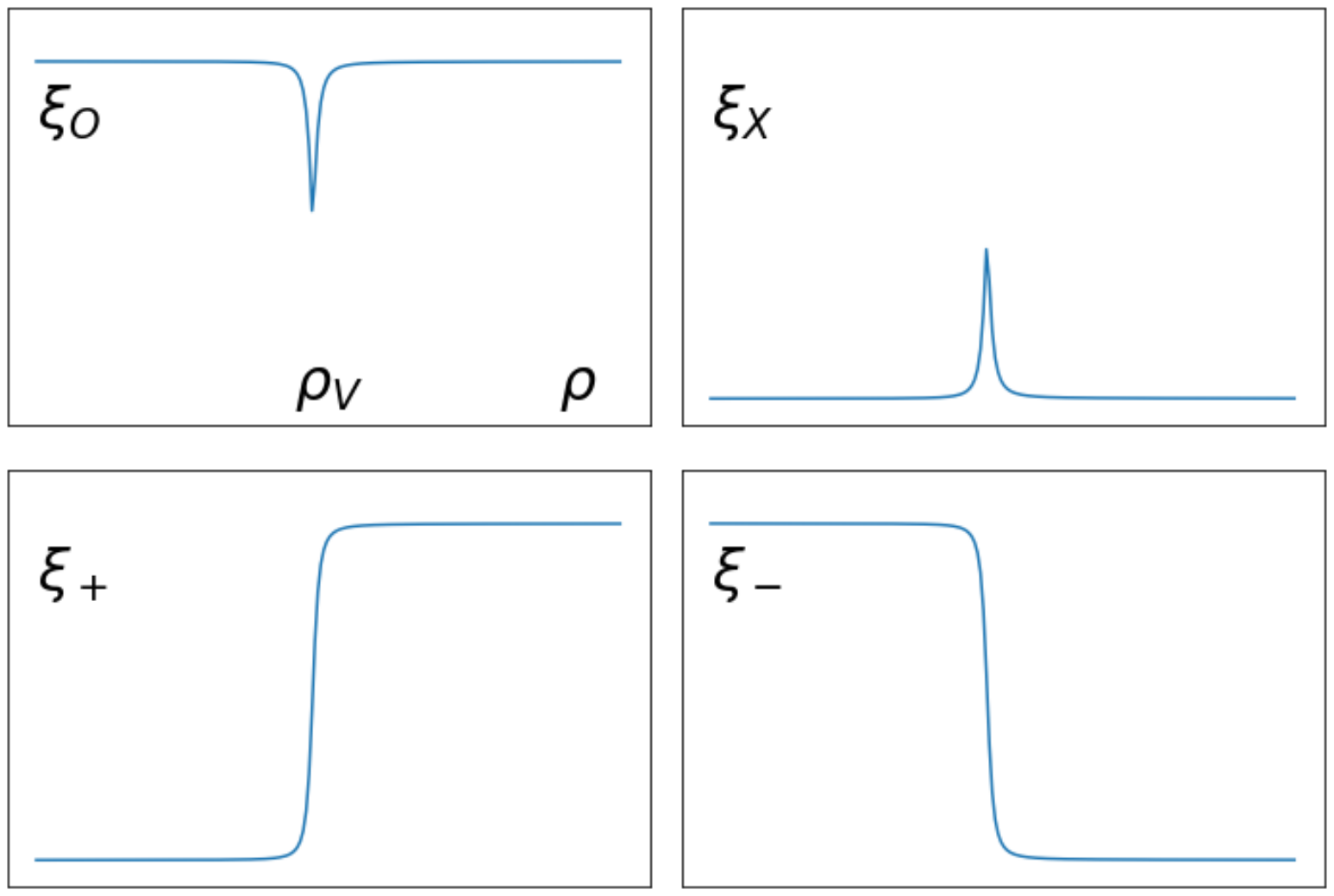}
\vskip -0.7cm
\caption{
The qualitative behavior of $\xi_i$ (the dimensionless mode opacity; see Eq.~\ref{eq:kappa}) as a function of density for photon mode-$i$ (with $i={\rm X}, {\rm O}, +, -$). At the vacuum resonance ($\rho=\rho_V$), $\xi_X$ has a spike, $\xi_O$ has a dip, and $\xi_+$ and $\xi_-$ have discontinuities.
Note that $\xi_+=\xi_X$ for $\rho<\rho_V$ and $\xi_+=\xi_O$ for $\rho>\rho_V$,
while $\xi_-=\xi_O$ for $\rho<\rho_V$ and $\xi_-=\xi_X$ for $\rho>\rho_V$.}
\label{fig:f2}
\end{figure}

For a given photon energy $E$ and wave vector ${\bf k}$ (which is inclined by angle
$\alpha$ with respect to the surface normal), 
the transfer equation for mode $i$ (with $i={\rm X}, {\rm O}$ or $i=+,\,-$) reads
\begin{equation}
\cos\alpha {\der I_i\over \der y} = \kappa_i \left(I_i-{1\over 2}B_\nu\right),
\label{eq:trans}\end{equation}
where $B_\nu(T)$ is the Planck function and $T=T(y)$ is the temperature profile (to be
specified later).
To calculate the emergent polarized radiation intensity 
from the atmosphere, taking account of partial mode conversion, we adopt the following procedure
\cite{vanadelsberg06}:
(i) We first integrate Eq.~(\ref{eq:trans}) for X-mode and O-mode 
($i={\rm X}, {\rm O}$) from large $y$ ($=\infty$) to $y_V$, the column density at which 
$\rho=\rho_V$. This gives $I_{\rm XV}$ and $I_{\rm OV}$, the X- and O-mode intensities just before
resonance crossing. (ii) We apply partial mode conversion
\begin{eqnarray}
&& I_{\rm XV}'= I_{\rm XV} P_J + I_{\rm OV} (1-P_J), \label{eq:Ixv'}  \\
&& I_{\rm OV}'= I_{\rm OV} P_J + I_{\rm XV} (1-P_J), \label{eq:Iov'}
\end{eqnarray}
to obtain the mode intensities just after resonance crossing. This partial conversion treatment is valid since the resonance width is small ($\Delta\rho/\rho_V\ll 1$; see Eq.~\ref{eq:width}).
(iii) We then integrate Eq.~(\ref{eq:trans}) for X-mode and O-mode again from $y=y_V$ (with the 
"initial" values $I_{\rm XV}', I_{\rm OV}'$) to $y\ll 1$.  This then gives the mode intensities emergent from the atmosphere, $I_{X,e}$, $I_{O,e}$.

An alternative procedure is to integrate Eq.~(\ref{eq:trans}) for $i={\rm X}, {\rm O}, +, -$ from 
$y\gg 1$ to $y\ll 1$ (without applying partial mode conversion), which gives the intensities $I_X(0), I_O(0), I_+(0)$ and $I_-(0)$ at $y\simeq 0$. Then apply the partial conversion
\begin{eqnarray}
&& I_{X,e}= I_X(0) P_J + I_{+}(0) (1-P_J), \label{eq:Ixe2}  \\
&& I_{O,e}= I_O(0) P_J + I_{-}(0) (1-P_J), \label{eq:Ioe2}
\end{eqnarray}
where $P_J$ is evaluated at $y=y_V$. 

It is straightforward to show that the above two procedures are equivalent. For example,
after obtaining $I_{\rm XV}'$, we can get the emergent X-mode intensity by
\begin{equation}
I_{X,e}=I_{\rm XV}'\exp\left(-{\tau_{\rm XV}\over\cos\alpha}\right) + \Delta I_{\rm XV},
\label{eq:Ixe}\end{equation}
where $\tau_{\rm XV}=\int_0^{y_V}\kappa_X\,\der y$ is the optical depth of X-mode (measured along the surface normal) at $y=y_V$, and $\Delta I_{\rm XV}$ is the contribution to $I_{X,e}$ from the region $0<y<y_V$:
\begin{equation}
\Delta I_{\rm XV}= \int_0^{\tau_{\rm XV}/\cos\alpha}\!\!\!\exp\left(-{\tau_X\over\cos\alpha}\right){1\over 2}B_\nu(T)\,{\der\tau_X\over\cos\alpha}.
\end{equation}
On the other hand, when integrating Eq.~(\ref{eq:trans}) for $i={\rm X}$ from $y\gg 1$ to $y\ll 1$,
we obtain
\begin{equation}
I_X(0)=I_{\rm XV}\exp\left(-{\tau_{\rm XV}\over\cos\alpha}\right) + \Delta I_{\rm XV}.
\label{eq:Ix0}\end{equation}
Similarly,
\begin{equation}
I_+(0)=I_{\rm OV}\exp\left(-{\tau_{\rm XV}\over\cos\alpha}\right) + \Delta I_{\rm XV}.
\label{eq:Ip0}
\end{equation}
It is easy to see that Eq.~(\ref{eq:Ixe}) together with Eq.~(\ref{eq:Ixv'}) (the first procedure)
and Eq.~(\ref{eq:Ixe2}) with Eqs.~(\ref{eq:Ix0})-(\ref{eq:Ip0}) (the second procedure) yield the same
emergent $I_{X,e}$.

\subsection*{Photosphere densities and Critical Field}

Before presenting our sample results, it is useful to estimate 
the photosphere densities for different modes and the condition
for polarization swing.

When the vacuum polarization effect is neglected ($\rho_V=0)$,
$|\beta|\gg 1$ at all densities (for typical $\theta_{\rm kB}$'s not too close to 0),
the $\xi$-factors for the O-mode ($|K|\gg 1$) and for the X-mode (with $|K|\ll 1$)
are
\begin{equation}
\xi_O\simeq \sin^2\theta_{\rm kB},\quad 
\xi_X\simeq {1\over u_e\sin^2\theta_{\rm kB}}.
\end{equation}
The photospheres of the O-mode and X-mode photons 
are determined by the condition
\begin{equation}
\int_0^{y_{O,X}}\! \kappa_{O,X}\, {{\rm d}y\over\cos\alpha}=2/3.
\end{equation}
The corresponding photosphere densities can be estimated as
\begin{eqnarray}
&& \rho_O = \xi_O^{-1/2}\rho_0,\\
&&\rho_X = \xi_X^{-1/2}\rho_0,
\end{eqnarray}
where the ``zero-field'' photosphere density is
\begin{equation}
\rho_0\simeq 0.59\,\left({\mu g_2\cos\alpha \over {\cal G}}\right)^{1/2}\!\left({E_1\over Z}\right)^{3/2}\!\biggl({A\over T_6^{1/4}}\biggr)\,\,{\rm g\,cm}^{-3}.
\end{equation}

\begin{figure}
\centering  
\includegraphics[height=7.8cm]{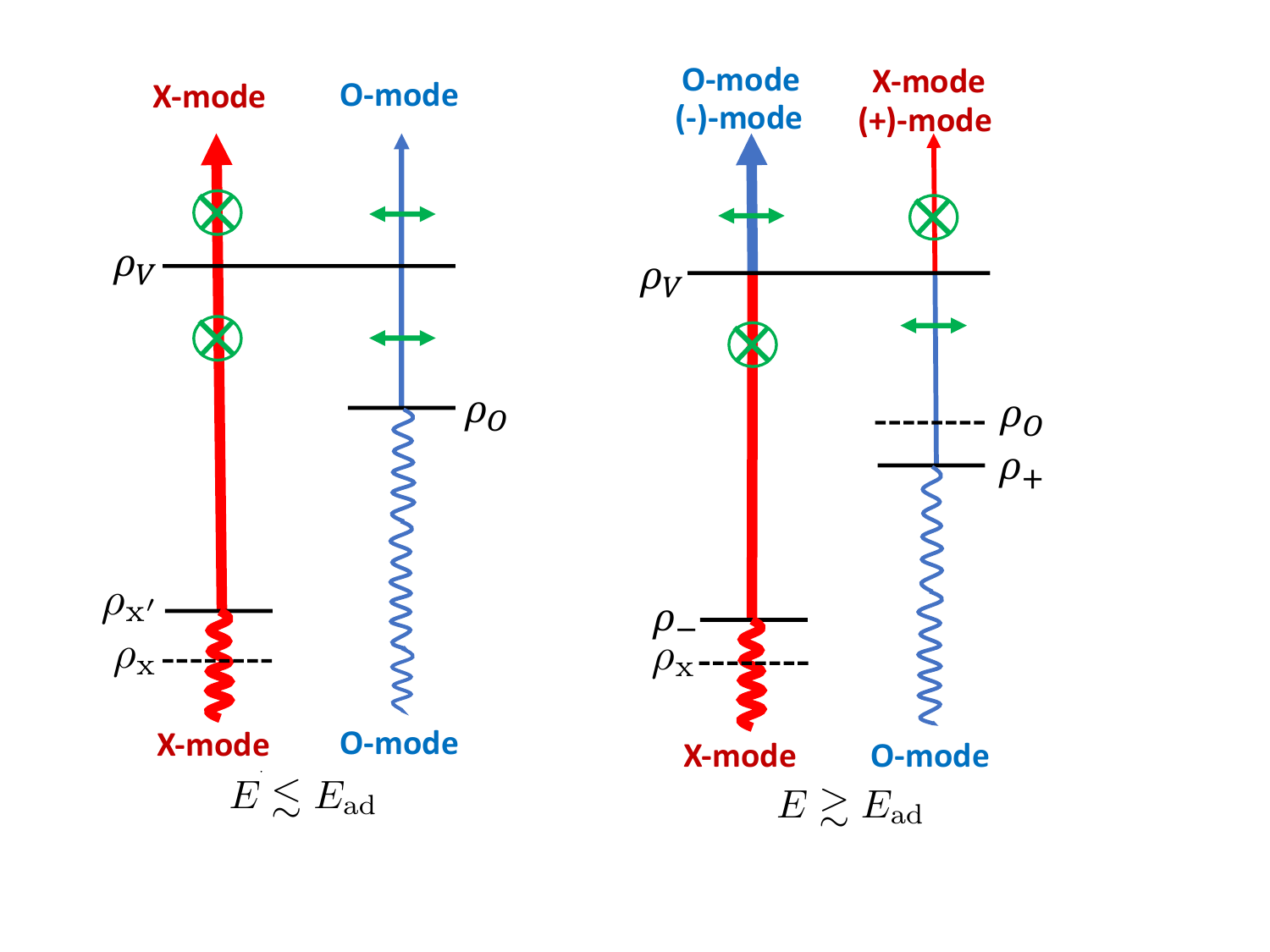}
\vskip -0.8cm
\caption{
A schematic diagram illustrating how vacuum resonance affects 
the polarization state of the emergent radiation from a magnetized 
NS atmosphere. This diagram applies to the $B\lesssim B_{\rm OV}$ regime
so that the vacuum resonance density $\rho_V$ is less than the O-mode
photosphere density $\rho_O$.
For $E\lesssim E_{\rm ad}$, the photon evolves nonadiabatically
across the vacuum resonance (for $\theta_{\rm kB}$ not too close to $0$), 
thus the emergent radiation is dominated by the X-mode. For 
$E\gtrsim E_{\rm ad}$, the photon evolves adiabatically, with its plane of
polarization rotating by 90$^\circ$ across the vacuum resonance, and thus
the emergent radiation is dominated by the O-mode. The plane of linear
polarization at low energies is therefore perpendicular to that at high 
energies.}
\label{fig:f3}
\end{figure}

\begin{figure}
\centering
\includegraphics[height=7.8cm]{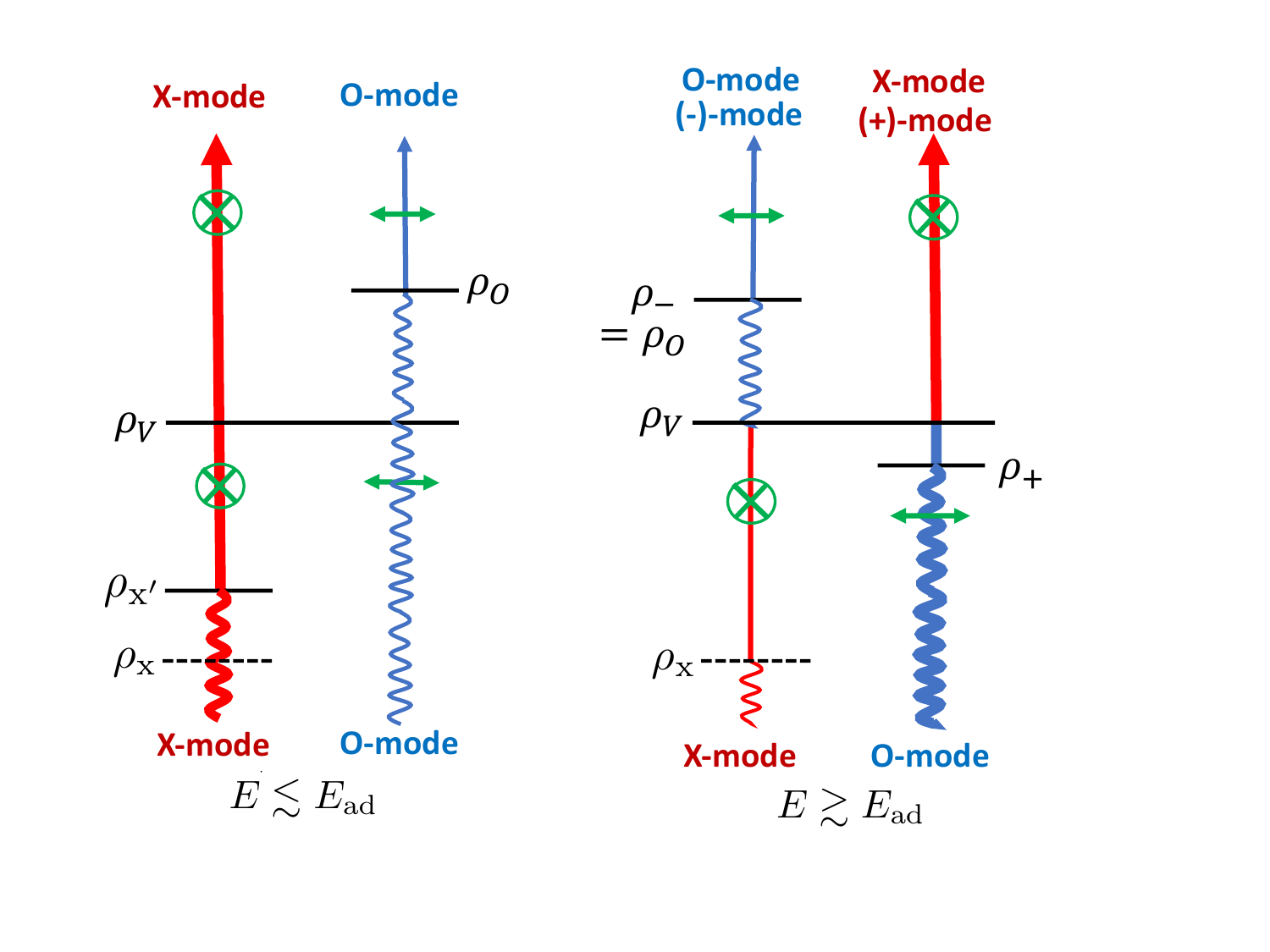}
\vskip -0.8cm
\caption{
Same as Fig.~\ref{fig:f3}, except for the $B\gtrsim B_{\rm OV}$ regime
($\rho_V>\rho_O$), in which 
the emergent radiation is always dominated by X-mode for all $E$'s.}
\label{fig:f4}
\end{figure}

The effect of the vacuum resonance on the radiative transfer 
depends qualitatively on the ratios $\rho_V/\rho_O$ and $\rho_V/\rho_X$, given by 
\begin{equation}
{\rho_V\over\rho_O}=\left({B\over B_{\rm OV}}\right)^2,\quad
{\rho_V\over\rho_X}={B\over B_{\rm XV}},
\end{equation}
where
\begin{eqnarray}
&&\!\!\! B_{\rm OV}=7.8\!\times\! 10^{13}\left(\!{\mu g_2\cos\alpha \over Z{\cal G}E_1\sin^2\!\theta_{\rm kB}}\!\right)^{\!1/4}\!\!\left(\!{f\over T_6^{1/8}}\!\right)
{\rm G},\label{eq:Bov}\\
&&\!\!\! B_{\rm XV}=7.1\!\times\! 10^{16}\left(\!{\mu g_2\cos\alpha\over Z{\cal G}E_1^3}\!\right)^{\!1/2}\!\!\left(\!{f^2\sin\theta_{\rm kB}\over T_6^{1/4}}\!\right) {\rm G}.
\end{eqnarray}
Clearly, the condition $B\ll B_{\rm XV}$ or $\rho_V\ll \rho_X$ is satisfied for almost all
relevant NS parameters of interest, while $B_{\rm OV}$ defines the
critical field strength for the 90$^\circ$ polarization swing (see Figs.~\ref{fig:f3}-\ref{fig:f4}): If $B\lesssim B_{\rm OV}$, the emergent radiation is dominated by 
the X-mode for $E\lesssim E_{\rm ad}$ and by the O-mode for
$E\gtrsim E_{\rm ad}$; if $B\gtrsim B_{\rm OV}$, the X-mode is dominant for all $E$'s.

We can quantify the role of $B_{\rm OV}$ more precisely by estimating
how vacuum resonance affects the photosphere densities.
In the limit of no mode conversion (i.e. $E\ll E_{\rm ad}$), it is appropriate to consider the transport of
X-mode and O-mode, with the mode opacities modified around 
the vacuum resonance (see Fig.~\ref{fig:f2}). The O-mode opacity has a dip 
near $\rho=\rho_V$ (where $\xi\simeq \sin^2\theta_{\rm kB}/2$), and since 
the resonance width $\Delta\rho/\rho_V\ll 1$, the photosphere density $\rho_{O'}$ 
is almost unchanged from the no-vacuum value, i.e.
$\rho_{O'}\simeq \rho_O$.
On the other hand, the X-mode opacity has a large spike at $\rho=\rho_V$
(where $\xi=\sin^2\theta_{\rm kB}/2$) compared to the
off-resonance value ($\xi\sim u_e^{-1}$). The X-mode optical depth
across the resonance (from $\rho_V-\Delta\rho$ to $\rho_V+\Delta\rho$)
is of order $\Delta\tau_V\sim \epsilon (\rho_V/\rho_O)^2$, where
$\epsilon$ is given by Eq.~(\ref{eq:width}). Thus the modified X-mode
photosphere density is $\rho_{X'}\simeq \rho_V$ for
$\Delta\tau_V\gtrsim 1$ and
\begin{equation}
\rho_{X'}\sim \rho_X \left[1-\epsilon  
\left({\rho_V/\rho_O}\right)^2\right]^{1/2}
\end{equation}
for $\Delta\tau_V\lesssim 1$.

In the limit of complete mode conversion (i.e. $E\gg E_{\rm ad}$), it
is appropriate to consider the transport of $(+)$-mode and $(-)$-mode,
with the mode opacities exhibiting a discontinuity at $\rho=\rho_V$
(see Fig.~2).  The $(+)$-mode photosphere density $\rho_+$ is given by
\begin{equation}
\rho_+^2\simeq \rho_O^2+\left(1+{\xi_X\over\xi_O}\right)\rho_V^2\simeq \rho_O^2+\rho_V^2.
\end{equation}
The $(-)$-mode photosphere density $\rho_-$ is only affected by the vacuum resonance if $\rho_O>\rho_V$. Thus
\begin{equation}
\rho_-=\rho_O  \qquad {\rm for}~~\rho_O<\rho_V 
\end{equation}
and 
\begin{equation}
\rho_-^2=\rho_V^2+{\xi_O\over\xi_X}\left(\rho_O^2-\rho_V^2\right) \quad {\rm for}~~\rho_O>\rho_V
\end{equation}

For general $E$'s with partial mode conversion, the emergent mode intensities are approximately given by
\begin{eqnarray}
&&I_{O,e}\simeq {1\over 2}P_J B_\nu(\rho_O) +{1\over 2}(1-P_J)B_\nu(\rho_-),\label{eq:I_Oe}\\
&&I_{X,e}\simeq {1\over 2}P_J B_\nu(\rho_{X'}) P_J +{1\over 2}(1-P_J) B_\nu(\rho_+),
\end{eqnarray}
where (for example) $B_\nu(\rho_O)$ is the Planck function evaluated at $\rho=\rho_O$.
These results are schematically depicted in Figs.~\ref{fig:f3}-\ref{fig:f4}.

\subsection*{Results}

To compute the polarized radiation spectrum emergent from a NS
atmosphere patch using the method presented above (Eq.~\ref{eq:trans} with
Eqs.~\ref{eq:Ixv'}-\ref{eq:Iov'} or with
Eqs.~\ref{eq:Ixe2}-\ref{eq:Ioe2}),
we need to know the atmosphere temperature profile $T(y)$. This can be
obtained only by self-consistent atmosphere modeling, which has only been
done for a small number of cases (in terms of the local $T_s$, ${\bf B}$ and composition).
Here, to explore the effect of different ${\bf B}$ and compositions, we consider two approximate models:

$\bullet$ Model (i): We use the profile $T_{\rm H}(y)$ for the
$T_s=5\times 10^6$~K H atmosphere model with a vertical field
$B=10^{14}$~G
[see Fig.~5 and Eq.~48 of
\cite{vanadelsberg06}; this model correctly treats
the partial model conversion effect], and re-scale it to take account
of the modification of the free-free opacity for different $Z,~A$ (so
that the re-scaled profile yields the same effective surface
temperature $T_s$):
\begin{equation}
    T(y)=\left({2\mu Z^3/ A^2}\right)^{1/8.5} T_{\rm H}(y).
\label{eq:rescale}\end{equation}

$\bullet$ Model (ii): We use a smooth (monotonic) fit to the $T_{\rm H}$ profile, given by 
\begin{equation}
\log_{10} T_{\rm H}(y)=0.11 +0.147\,\left[\,\log_{10}(0.4y)+3\,\right],
\end{equation}
and then apply Eq.~(\ref{eq:rescale}) for re-scaling.

\begin{figure}
\centering
\includegraphics[height=10.2cm]{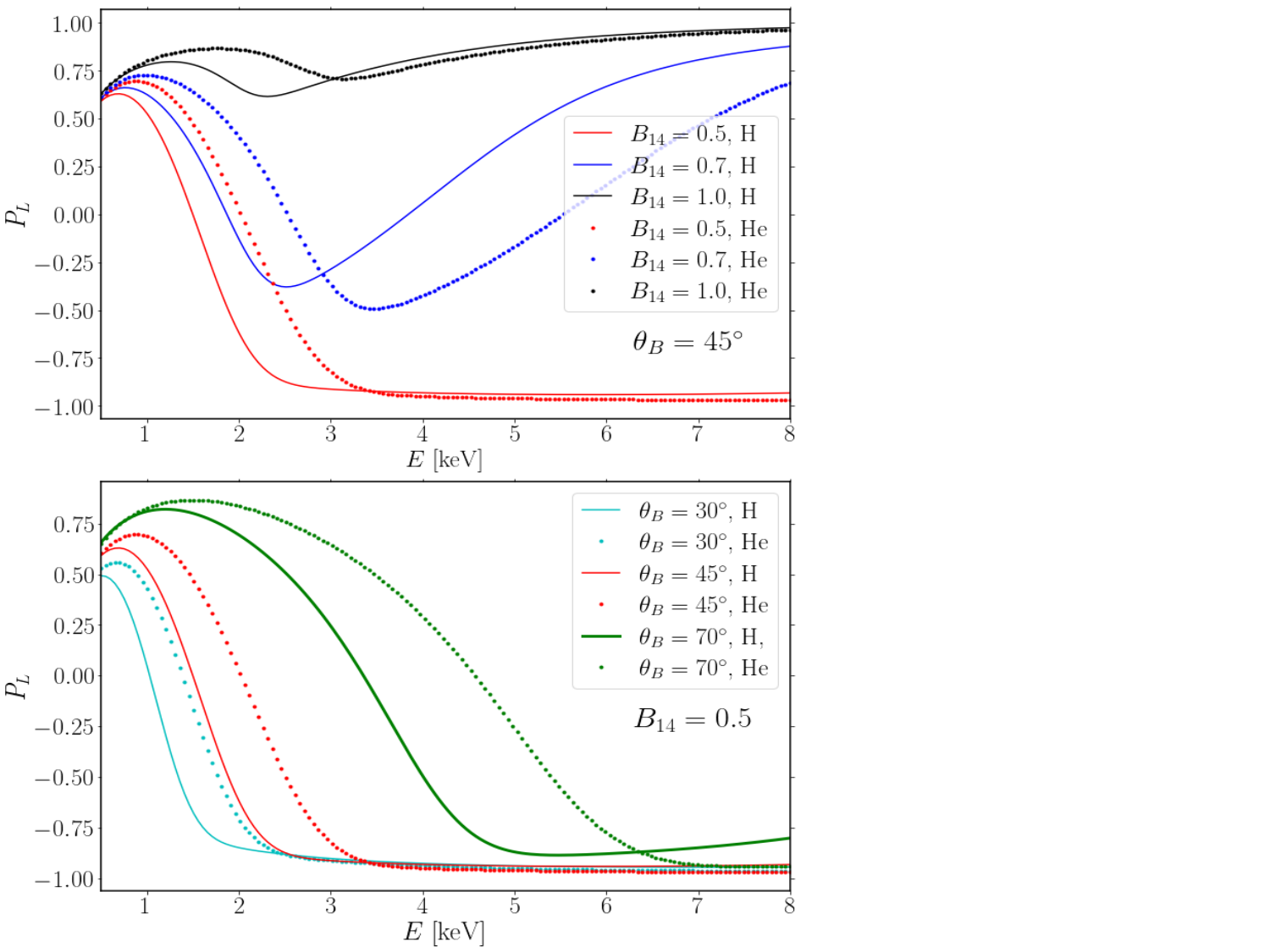}
\caption{
  Polarization degree $P_L$ (defined by Eq.~\ref{eq:pl}) of
  the emergent radiation normal to the surface as a function of the photon energy $E$ for H
  and He atmospheres with different magnetic field strengths and
  directions ($\theta_B$, the angle between the surface ${\bf B}$ and
  the surface normal vector). All results are based on the temperature
  profile Model (ii).}
\label{fig:f5}
\end{figure}

\begin{figure}
\centering
\includegraphics[height=10.3cm]{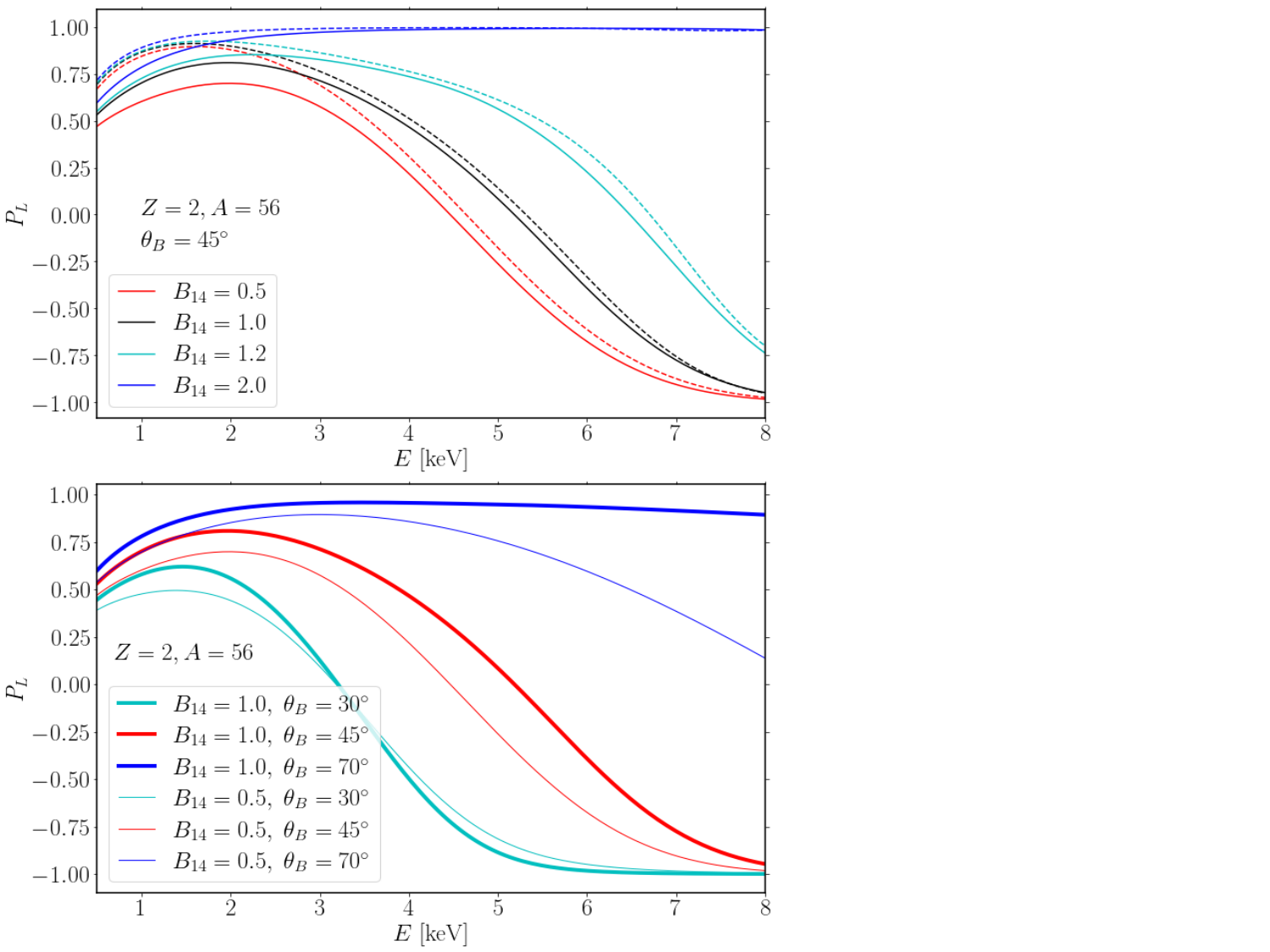}
\caption{
  Similar to Fig.~\ref{fig:f5}, except for partially ionized Fe atmospheres with $Z=2,~A=56$.
  In the upper panel, the solid/dashed lines are for Model (i)/(ii). In the lower panel, the results are shown 
  for Model (i).}
\label{fig:f6}
\end{figure}

Figures \ref{fig:f5}-\ref{fig:f6} show a sample of our results for the
polarization degree of the emergent radiation, defined by
\be
P_L\equiv {I_{X,e}-I_{O,e}\over I_{X,e}+I_{O,e}}.
\label{eq:pl}\ee
We see that for the H and He atmospheres (Fig.~\ref{fig:f5}), $P_L$
transitions from being positive at low $E$'s to negative at high $E$'s
for $B_{14}\lesssim 0.5$, in agreement with the critical field
estimate (Eq.~\ref{eq:Bov}). The transition energy (where $P_L=0$) is
approximately given by $E_{\rm ad}$, and has the scaling $E_{\rm ad}
\propto (\mu\tan^2\theta_B)^{1/3}$, where $\theta_B$ is the angle between
the surface ${\bf B}$ and the surface normal (see Eq.~\ref{condition}). To
obtain a transition energy of $4-5$~keV (as observed for 4U 0142)
would require most of the emission comes from the surface region with
$\theta_B\gtrsim 70^\circ$.

On the other hand, for a partially ionized heavy-element atmosphere
(such that $\mu/Z$ is much larger than unity), the critical field
$B_{\rm OV}$ can be increased (see Eq.~\ref{eq:Bov}). Figure~\ref{fig:f6}
shows that at $B_{14}=1$, a $Z=2,~A=56$ atmosphere can have a sign
change in $P_L$ around $E\sim 3-5$~keV (depending on the $\theta_B$
value.

To determine the observed polarization signal, we must consider the propagation
of polarized radiation in the NS magnetosphere, whose dielectric
property in the X-ray band is dominated by vacuum polarization
\cite{heyletal03}.
As a photon propagates from the
NS surface through the magnetosphere, its polarization state evolves
following the varying magnetic field it experiences, up
to the ``polarization-limiting radius'' $r_{\rm pl}$, beyond which the
polarization state is frozen. It is convenient to set up
a fixed coordinate system $XYZ$, where the Z-axis is
along the line-of-sight and the $X$-axis lies in the plane
spanned by the $Z$-axis and ${\bf\Omega}$ (the NS spin angular velocity
vector). The polarization-limiting radius $r_{\rm pl}$ is determined by the condition
$\Delta k=2|d\phi_B/ds|$, where $\Delta k=|k_X-k_O|$ is the difference in the wavenumbers of the
two photon modes, and $\phi_B(s)$ is the azimuthal angle of the magnetic field along the
ray ($s$ measures the distance along the ray). For a NS with surface dipole field $B_d$ and spin
frequency $\nu=\Omega/(2\pi)$, we have
\cite{vanadelsberg06}
${r_{\rm pl}/R}\sim 150\, ( E_1 B_{d,14}^2/\nu_1 )^{1/6}$, where 
$B_{d,14}=B_d/(10^{14}\,{\rm G})$ and $\nu_1=\nu/{\rm Hz}$. Note that since
$R\ll r_{\rm pl}\ll r_l$ [with $r_l=c/\Omega$ the light-cylinder radius),
only the dipole field determines $r_{\rm pl}$.
Regardless of the surface magnetic field structure, the radiation emerging from most atmosphere patches
with mode intensities $I_{X,e}$ and $I_{O,e}$ evolves adiabatically in the magnetosphere such that
the radiation at $r> r_{\rm pl}$ consists of approximately the same $I_{X,e}$ amd $I_{O,e}$,
with a small mixture of circular polarization generated around $r_{\rm pl}$
\cite{vanadelsberg06}.
The exception occurs for those rays that encounter the quasi-tangential point
(where the photon momentum is nearly aligned with the local magnetic field)
during their travel from the surface to $r_{\rm pl}$
\cite{wang09}.
Since only a small fraction of the NS surface radiation is affected by the quasi-tangential
propagation, we can neglect its effect if the observed radiation comes from a large area of the NS surface.
Let the azimuthal angle of the ${\bf B}$ field at $r_{\rm pl}$ be $\phi_B(r_{\rm pl})$ in the
the $XYZ$ coordinate system. The observed
Stokes parameters (normalized to the total intensity $I$) are then given by
\begin{eqnarray}
&& Q/I=-P_L \cos 2\phi_B(r_{\rm pl}),\\
&& U/I=-P_L\sin 2\phi_B(r_{\rm pl}).
\end{eqnarray}
Note that when $r_{\rm pl}\ll r_l$, the transverse ($XY$) component of ${\bf B}(r_{\rm pl})$
is opposite to the transverse component of the magnetic dipole moment ${\boldsymbol \mu}$, thus
$\phi_B(r_{\rm pl})\simeq \pi +\phi_\mu$, where $\phi_\mu$ is the azimuthal angle of 
${\boldsymbol \mu}$.

For determine the emission from the whole NS surface, we need to add
up contributions from different patches (area element $\der S$),
inclduing the effect of general relativity
\cite{pechenick83,beloborodov02}.
For example, the observed spectral fluxes $F_I,~F_Q$ (associated with the intensities $I,~Q$) are
\ba
&& F_I= g_r^3 \int\! {\der S\,\cos\alpha\over D^2}(I_{X,e}+I_{O,e}),\\
&& F_Q= g_r^3 \int\! {\der S\,\cos\alpha\over D^2}(I_{O,e}-I_{X,e})\cos 2\phi_B(r_{\rm pl}),
\ea
where $g_r\equiv (1-2GM/Rc^2)^{1/2}$ and $\alpha$ is the angle between
the ray and the surface normal at the emission point). Clearly, to
compute $F_I,~F_Q$ and $F_U$ requires the knowledge of the
distributions of NS surface temperature and magnetic field, as well as various
angles (the relative orientations between the line of sight, the spin axis and the dipole axis).
This is beyond the scope of the paper.
[Note that since $\phi_B(r_{\rm pl})\simeq \pi +\phi_\mu$,
the phase variation of $F_Q,\,F_U$ follows the rotation of the magnetic dipole, as in the
rotating vector model.]
Nevertheless, our results depicted
in Figs.~\ref{fig:f5}-\ref{fig:f6} (with different values of local field strengths and orientations)
show that the 90$^\circ$ linear polarization swing observed in AXP 4U 0142 can be explained
by emission from a partially ionized heavy-element atmosphere with surface field strength
about $10^{14}$~G, or a H/He atmosphere with $B_{14}\lesssim 0.5$ and a more restricted surface
field geometry (i.e. most of the radiation comes from the regions with $\theta_B\gtrsim 70^\circ$).

\section*{Discussion}

We have shown that the observed the x-ray polarization signal from AXP
4U 0142, particularly the 90$^\circ$ swing around 4-5~keV, can be
naturally explained by the mode conversion effect associated vacuum
resonance in the NS atmosphere. In this scenario, the 2-4~keV emission
is dominated by the X-mode, while the 5-8~keV emission by the O-mode
as a result of the adiabatic mode conversion from the X-mode to the
O-mode.
This interpretation of the polarization swing would imply that the
NS's kick velocity is aligned with its spin axis, in agreement with
the spin-kick correlation observed in other NS systems (see
``Introduction'').

It is important to note that in our scenario, the existence of the
X-ray polarization swing depends sensitively on the actual value of
the magnetic field on the NS surface (see Eq.~\ref{eq:Bov}). To
explain the observation of AXP 4U 0142, the magnetic fields in most
region of the NS surface must be less than about $10^{14}$~G, and a
lower field strength would be preferred in terms of producing the
polarization swing robustly (for a wide range of geometrical parameters).
Using the pulsar spindown power derived from force-free electrodynamics
simulations
\cite{spitkovsky06},
$L_{\rm sd}=(\mu^2\Omega^4/c^3)(1+\sin^2\theta_{s\mu})$
(where $\theta_{s\mu}$ is the angle between the magnetic dipole ${\boldsymbol \mu}$ and the spin axis),
we find that for the observed $P,\,\dot P$ of 4U 0142,
the dipole field at the magnetic equator is
$B_d=1.1\times 10^{14}I_{45}^{1/2}R_{6}^{-3}(1+\sin^2\theta_{s\mu})^{-1/2}$~G, where
$R_6$ is the NS radius in units of $10^6$~cm, and $I_{45}$ is the moment of inertia in units of
$10^{45}$~g~cm$^2$. Note that with $R_6\simeq 1.3$
\cite{miller21,raaijmakers21},
the above estimate is reduced by a factor of 2.  In addition, if the magnetar
possesses a relativistc wind with luminosity $L_w\gtrsim L_{\rm sd}$, the wind can open up field
lines at $r_{\rm open}\sim (B_d^2R^6c/L_w)^{1/4}\lesssim c/\Omega$ and significantly enhance
the spindown torque
\cite{thompson98,harding99,thompson00}.
This would imply that a smaller $B_d$ is needed to produce the observed $\dot P$ in 4U 0142.
Thus, the observation of X-ray polarization swing is consistent with
the indirectly ``measured'' dipole field, and requires that the
high-order multipole field components be not much stronger than the
dipole field.

We reiterate some of the caveats of our work: we have not
attempted to calculate the synthetic polarized radiation from the
whole NS and to compare with the X-ray data from AXP 4U 0142 in
detail; our treatment of partially ionized heavy-element magnetic NS
atmospheres is also approximate in several aspects (e.g., bound-free
opacities are neglected, the vertical temperature profiles are assumed
based on limited H atmosphere models). At present, these caveats are
unavoidable, given the uncertainties (and large parameter space needed
to do a proper survey) in the surface temperature and magnetic field
distributions on the NS, and the fact that self-consistent
heavy-element atmospheres models for general magnetic field strengths
and orientations have not be constructed, especially in the regime
where the vacuum resonance effects are important.

In this work, we have interpreted the 2-8~keV polarization signal from
AXP 4U 0142 in terms of thermal emission from the NS surface.  It is
well established that for quiescent magnetars, the spectrum turns up
above 10 keV, such that the bulk of their energy comes out as
non-thermal hard X-rays
\cite{kuiper04,kaspi17}.
The 0.5-10~keV spectrum can be parameterized either by an absorbed
blackbody plus a power-law component (of photon index between $-4$ and
$-2$) or by the sum of several blackbodies.  So it is possible that
the 5-8 keV emission from AXP 4U 0142 has a significant non-thermal
contribution.
Thompson \& Kostenko \cite{thompson20}
have studied a model in
which the hard X-ray emission of quiescent magnetars comes from the
``annihilation bremsstrahlung'' of an electron–positron magnetospheric
plasma. Such emission is mainly polarized in O-mode. If this emission
dominates the 5-8~keV spectrum of APX 4U 0142 while negligible at
lower energies, it could provide an alternative explanation for the
polarization swing for a wide range of surface field strengths
(recall that for $B\gtrsim B_{\rm OV}$, the atmosphere thermal emission is
dominated by X-mode for all $E$'s (see eq.~\ref{eq:Bov}).

Very recently, while this paper is under review, IXPE reported the detection
of polarized X-rays from another magnetar, 1RXS J170849.0-400910 \cite{zane23}:
The phase-averaged polarization signal exhibits 
an increase with energy, from $\sim 20\%$ at 2–3 keV to $\sim 80\%$ at 6–8 keV,
while the polarization angle is independent of the photon energy.
This constant polarization angle is consistent with atmospheric emission dominated by X-mode at all energies,
and is expected since the measured dipole field (based on spindown) for this AXP,
$B_d\simeq 5\times 10^{14}$~G, is significantly larger than $B_{\rm OV}$
(eq.~\ref{eq:Bov}). The increase of linear polarization with $E$ can be a result of the magnetic
field geometry (relative to the spin axis and line-of-sight) and the surface temperature
distribution [see Ref.~\cite{vanadelsberg06} for examples].

Overall, our work demonstrates the important role played by the vacuum
resonance in producing the observed X-ray polarization signature from magnetars
(and NSs with weaker magnetic fields).
The observations of AXP 4U 0142 and 1RXS J170849.0-400910
by IXPE have now opened up a
new window in studying the surface environment of NSs.  Future X-ray
polarization mission [such as eXTP;
see Ref.~\cite{zhang19}]
will provide more detailed observational data. Comprehensive
theoretical modelings of magnetic NS surface radiation and
magnetosphere emission will be needed to confront these observations.

\noindent
{\bf Acknowledgements:}
We thank Chris Thompson for useful discussion.
This work is supported in part by Cornell University.



\begin{thebibliography}{99}

\bibitem[Thompson \& Duncan 1993]{thompsonduncan93}
  Thompson, C., Duncan R. C. 1993,
Neutron star dynamos and the origins of pulsar magnetism,
  ApJ, 408, 194

\bibitem[Kaspi \& Beloborodov 2017]{kaspi17}
  Kaspi V. M., Beloborodov A. M., 2017,
  Magnetars,
  ARA\&A, 55, 261                                                       

\bibitem[Weisskopf et al.~2016]{weisskopf16}
  Weisskopf, M.C. et al. 2016,
The imaging x-ray polarimetry explorer (IXPE),
  Proc. SPIE 9905, 990517; https://doi.org/10.1117/12.2235240                    

\bibitem[Taverna et al.~2022]{taverna22}
  Taverna, R. et al. 2022,
  Polarized x-rays from a magnetar,
  Science, 378, 646

\bibitem[Medin \& Lai 2007]{medin07}
Medin, Z., Lai, D. 2007,
  Condensed surfaces of magnetic neutron stars, thermal surface emission, and particle acceleration above pulsar polar caps,
  MNRAS, 382, 1833                                                                   

\bibitem[Potekhin \& Chabrier 2013]{potekhin13}
  Potekhin, A., Chabrier, G. 2013,
Equation of state for magnetized Coulomb plasmas,
  A\&A, 550, A43                                                             

\bibitem[Medin \& Lai 2006]{medin06}
  Medin, Z., Lai, D. 2006,
Density-functional-theory calculations of matter in strong magnetic fields. II. Infinite chains and condensed matter,
  Phy. Rev. A 74, 062508                                                             

\bibitem[van Adelsberg et al.~2005]{vanadelsberg05}
  van Adelsberg, M., Lai, D. Potekhin, A.Y., Arras, P. 2005,
  Radiation from condensed surface of magnetic neutron stars,
  ApJ, 628, 902                                    

\bibitem[Potekhin et al.~2012]{potekhin12}
  Potekhin, A.Y., Suleimanov, V.F., van Adelsberg, M., Werner, K. 2012,
  Radiative properties of magnetic neutron stars with metallic surfaces and thin atmospheres,
  A\&A, 546, A121                       

\bibitem[Lai et al.~2001]{lai01}
  Lai, D., Chernoff, D.F., Cordes, J.M. 2001,
Pulsar jets: implications for neutron star kicks and initial spins,
  ApJ, 549, 1118                                                  

\bibitem[Johnston et al.~2005]{johnston05}
  Johnston, S., et al. 2005,
Evidence for alignment of the rotation and velocity vectors in pulsars,
  MNRAS, 364, 1394                                                                 

\bibitem[Wang et al.~2006]{wang06}
  Wang, C., Lai, D., Han, J.L. 2006,
  Neutron Star Kicks in Isolated and Binary Pulsars: Observational
  Constraints and Implications for Kick Mechanisms,
  ApJ, 639, 1007

\bibitem[Noutsos et al.~2013]{noutsos13}
  Noutsos, A., et al. 2013,
  Pulsar spin–velocity alignment: kinematic ages, birth periods and braking indices,
  MNRAS, 430, 2281                                                                  
  
\bibitem[Janka et al.~2022]{janka22}
Janka, H.-T., Wongwathanarat, A., Kramer, M. 2022,                                                          
Supernova fallback as origin of neutron star spins and spin-kick alignment,
ApJ, 926, 9 

\bibitem[Lai \& Ho 2003a]{laiho03a}
  Lai, D., Ho, W.C.G. 2003a,
  Polarized X-ray emission from magnetized neutron stars: signature of strong-field vacuum polarization,
  PRL, 91, 071101                                                                  
  
\bibitem[Zane et al.~2023]{zane23}
  Zane, S., et al.~2023,
A strong X-ray polarization signal from the magnetar 1RXS J170849.0-400910,
  ApJ, arXiv:2301.12919
  
\bibitem[Heisenberg \& Euler 1936]{heisenbergeuler36} 
Heisenberg, W. \& Euler, H. 1936, 
Folgerungen aus der Diracschen Theorie des Positrons,
Z. Physik, 98, 714

\bibitem[Schwinger 1951]{schwinger51} 
Schwinger, J. 1951, 
On Gauge Invariance and Vacuum Polarization,
Phys. Rev., 82, 664

\bibitem[Adler 1971]{adler71}
Adler, S.L. 1971, 
Photon Splitting and Photon Dispersion in a Strong Magnetic Field,
Ann. Phys., 67, 599

\bibitem[Tsai \& Erber 1975]{tsaierber75} 
Tsai, W.Y.,  Erber, T. 1975, 
Propagation of photons in homogeneous magnetic fields: Index of refraction,
Phys. Rev. D, 12, 1132

\bibitem[Heyl \& Hernquist 1997]{heylhernquist97} 
Heyl, J.S. \& Hernquist, L. 1997, 
Birefringence and dichroism of the QED vacuum,
J. Phys. A 30, 6485

\bibitem[Gnedin et al.~1978]{gnedinetal78} 
Gnedin, Yu.N., Pavlov, G.G., \& Shibanov, Yu.A. 1978, 
The Effect of Vacuum Birefringence in a Magnetic Field on the Polarization 
and Beaming of X-ray Pulsars,
Sov. Astron. Lett., 4(3), 117

\bibitem[M\'{e}sz\'{a}ros \& Ventura 1979]{meszarosventura79}
M\'{e}sz\'{a}ros, P. \& Ventura, J. 1979, 
Vacuum polarization effects on radiative opacities in a strong magnetic field,
Phys. Rev. D 19, 3565

\bibitem[Pavlov \& Gnedin 1984]{pavlovgnedin84} 
Pavlov, G.G. \& Gnedin, Yu.N. 1984, 
Vacuum Polarization by a Magnetic Field and its Astrophysical Manifestations,
Sov. Sci. Rev. E: Astrophys. Space Phys. 3, 197

\bibitem[Lai \& Ho 2022]{laiho02}
Lai, D., Ho, W.C.G. 2002,
Resonant Conversion of Photon Modes Due to Vacuum
Polarization in a Magnetized Plasma: Implications for X-Ray Emission from
Magnetars,
ApJ, 566, 373

\bibitem[Ho \& Lai 2003]{holai03}
Ho, W.C.G., \& Lai, D. 2003, 
Atmospheres and Spectra of Strongly Magnetized Neutron Stars II: Effect of
Vacuum Polarization,
MNRAS, 338, 233

\bibitem[Lai \& Ho 2003b]{laiho03b}
Lai, D., Ho, W.C.G. 2003b, 
Transfer of Polarized Radiation in Strongly Magnetized Plasmas and Thermal Emission from Magnetars: Effect of Vacuum Polarization,
ApJ, 588, 962

\bibitem[M\'{e}sz\'{a}ros 1992]{meszaros92}
M\'{e}sz\'{a}ros, P. 1992,
High-Energy Radiation from Magnetized Neutron Stars
(Univ. Chicago Press, Chicago)

\bibitem[Potekhin et al.~2004]{potekhin04}
  Potekhin, A., et al.~2004,
Electromagnetic Polarization in Partially Ionized Plasmas with Strong Magnetic Fields and Neutron Star Atmosphere Models,
  ApJ, 612, 1034                                                                   

\bibitem[Haxton 1995]{haxton95} 
Haxton, W.C. 1995, 
 The Solar Neutrino Problem,
Ann. Rev. Astron. Astrophys., 33, 459

\bibitem[Bahcall et al.~2003]{bahcall03}
Bahcall, J.N., Gonzalez-Garcia, M.C., \& Pena-Garay, C. 2003,
Solar Neutrinos Before and After KamLAND,
J. High Energy Phys. JHEP02, 009

\bibitem[Tracy et al.~2014]{tracy14}
Tracy, E.R., et al. 2014, Ray Tracing and Beyond: 
Phase Space Methods in Plasma Wave Theory (Cambridge Univ.)

\bibitem[van Adelsberg \& Lai 2006]{vanadelsberg06}
  van Adelsberg, M., Lai, D. 2006,
  Atmosphere models of magnetized neutron stars: QED effects, radiation spectra and polarization signals,
  MNRAS, 373, 1495                                                           

\bibitem[Zane 2006]{zane06}
  Zane, S., \& Turolla, R. 2006,
  Unveiling the thermal and magnetic map of neutron star surfaces
  though their X-ray emission: method and light-curve analysis,
  MNRAS, 366, 727                                                               

\bibitem[Shabaltas \& Lai 2012]{shabaltas12}
  Shabaltas, N., Lai, D. 2012,
  The hidden magnetic fields of the young neutron star in Kes 79,
  ApJ, 748, 148
  
\bibitem[Taverna et al.~2020]{taverna20}
  Taverna, R., Turolla, R., Suleimanov, V., Potekhin, A.Y., Zane, S. 2020,
X-ray spectra and polarization from magnetar candidates,
  MNRAS, 492, 5057                   

\bibitem[Caiazzon et al.~2022]{caiazzo22}
  Caiazzo, I., Gonzalez-Caniulef, D., Heyl, J., Fernandez, R. 2022,
Probing magnetar emission mechanisms with X-ray spectropolarimetry,
  MNRAS, 514, 5024                          

\bibitem[Ho \& Lai 2002]{holai01}
Ho, W.C.G. \& Lai, D. 2001, 
Atmospheres and spectra of strongly magnetized neutron stars,
MNRAS, 327, 1081

\bibitem[Heyl \& Shaviv 2003]{heyletal03}
Heyl, J.S., Shaviv, N.J., \& Lloyd, D. 2003,
The High-Energy Polarization-Limiting Radius of Neutron Star Magnetospheres I
-- Slowly Rotating Neutron Stars,
MNRAS, 342, 134
  
\bibitem[Wang \& Lai 2009]{wang09}
  Wang, C., Lai, D. 2009,
  Polarization evolution in a strongly magnetized vacuum: QED effect
  and polarized X-ray emission from magnetized neutron stars,
  MNRAS, 398, 515

\bibitem[Pechenick et al.~1983]{pechenick83}
Pechenick, K.R., Ftaclas, C., \& Cohen, J.M. 1983,
Hot Spots on Neutron Stars: The Near-Field Gravitational Lens,
ApJ, 274, 846

\bibitem[Beloborodov 2002]{beloborodov02}
Beloborodov, A.M. 2002, 
Gravitational Bending of Light Near Compact Objects,
ApJ, 566, L85

\bibitem[Spitkovsky 2006]{spitkovsky06}
  Spitkovsky, A. 2006,
  Time-dependent Force-free Pulsar Magnetospheres: Axisymmetric and Oblique Rotators,
  ApJ, 648, L51                                                                          
  
\bibitem[Miller et al.~2021]{miller21}
  Miller, M.C. et al. 2021,
  The radius of PSR J0740+ 6620 from NICER and XMM-Newton data,
  ApJ, 918, L28

\bibitem[Raaijmakers et al.~2021]{raaijmakers21}
  Raaijmakers, G. et al. 2021,
  Constraints on the Dense Matter Equation of State and Neutron Star Properties from NICER's Mass–Radius Estimate of PSR J0740+6620 and Multimessenger Observations,
  ApJ, 918, L29
  
\bibitem[Thompson \& Blaes 1998]{thompson98}
  Thompson C., Blaes O., 1998,
Magnetohydrodynamics in the extreme relativistic limit,
  Phys. Rev. D 57, 3219                                                          
  
\bibitem[Harding et al.~1999]{harding99}
  Harding A. K., Contopoulos I., Kazanas D., 1999,
Magnetar spin-down,
  ApJ, 525, L125                                             
  
\bibitem[Thompson et al.~2000]{thompson00}
  Thompson C., Duncan R. C., Woods P. M., Kouveliotou C., Finger M. H., van Paradijs J., 2000,
Physical mechanisms for the variable spin-down and light curve of SGR 1900+14,
  ApJ, 543, 340

\bibitem[Kuiper et al.~2004]{kuiper04}
  Kuiper L., Hermsen W., Mendez M. 2004,
  Discovery of hard nonthermal pulsed X-ray emission from the anomalous X-ray pulsar 1E 1841–045,
  ApJ, 613, 1173
  
\bibitem[Thompson \& Kostenko 2020]{thompson20}
  Thompson C., Kostenko A. 2020,
Pair plasma in super-QED magnetic fields and the hard X-ray/optical emission of magnetars,
  ApJ, 904, 184

\bibitem[Zhang et al.~2019]{zhang19}
  Zhang, Shuangnan, et al. 2019,
   The enhanced X-ray Timing and Polarimetry mission—eXTP,
  Science China Physics, Mechanics \& Astronomy, 62, 29502                      




\end{thebibliography}
\end{document}